\documentclass[cup5b]{caps}
\usepackage{graphicx}
\usepackage{amssymb}
\usepackage{ociwsymp4e}  

\HeadText{T. Bensby, S. Feltzing, \& I. Lundstr\"om}  

\begin{document}

\pagenumbering{arabic}

\author[]{T. BENSBY, S. FELTZING, and I. LUNDSTR\"OM \\Lund  Observatory, Lund University, Sweden}

\chapter{A differential study of \\ the oxygen abundances in the \\ Galactic thin and thick disks}

\begin{abstract}

First results from a study into the abundance trends of oxygen in the Galactic thin and thick
disks are presented. Oxygen abundances for 63 nearby F and G dwarf stars,
based on very high resolution spectra ($R\sim215\,000$) and high signal-to-noise ($S/N>400$)
of the faint forbidden oxygen line at 6300 {\AA}, have been determined. 
Our findings can be summarized as follows: 
{\bf 1)} at $\rm [Fe/H]<0$ the oxygen trends in the thin and thick disk are smooth and distinct, 
indicating their different origins,
{\bf 2)} $\rm [O/Fe]$ for the thick disk stars show a turn-over at [Fe/H]$\sim -0.35$, indicating 
the peak of the enrichment from SNe type Ia to the interstellar medium,
{\bf 3)} the thin disk stars show a shallow decrease going from $\rm [Fe/H] \sim -0.7$ to the 
highest metallicities with no apparent turn-over present indicating a more quiet star formation history,
and {\bf 4)} $\rm [O/Fe]$ continues to decrease at $\rm [Fe/H]>0$ without showing
the leveling out that previously has been seen.

\end{abstract}

  \begin{figure}
    \centering
    \includegraphics[width=10cm,angle=0]{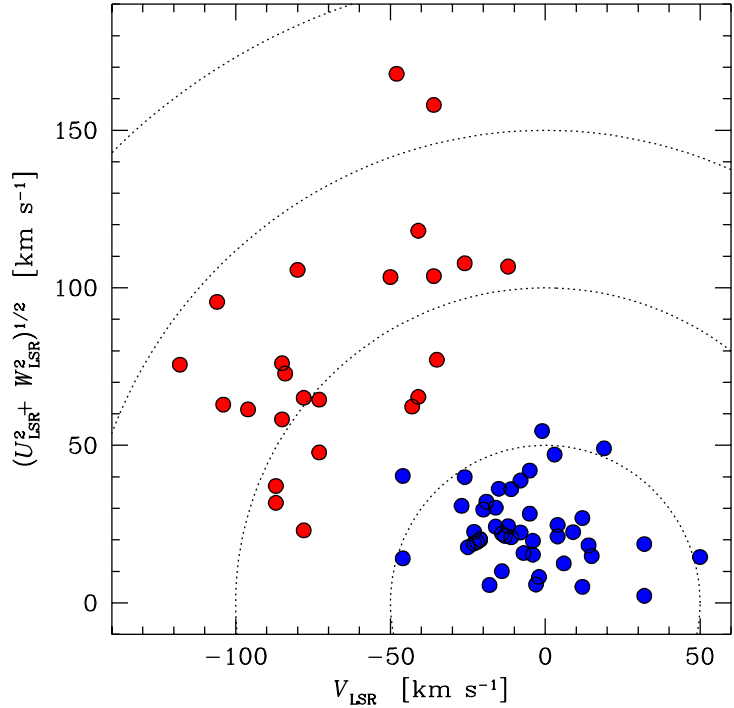}
    \caption{Kinematical data for the selected stars. Dash--dotted lines indicate constant peculiar
         space velocities $v_{\rm pec} = (U^{2} + V^{2} + W^{2})^{1/2}$ in steps of 50 km/s.
         Thick disk stars are marked as red, and thin disk stars as blue.}
    \label{fig:toomre}
  \end{figure}
  \begin{figure}
    \centering
    \includegraphics[width=10cm,angle=0]{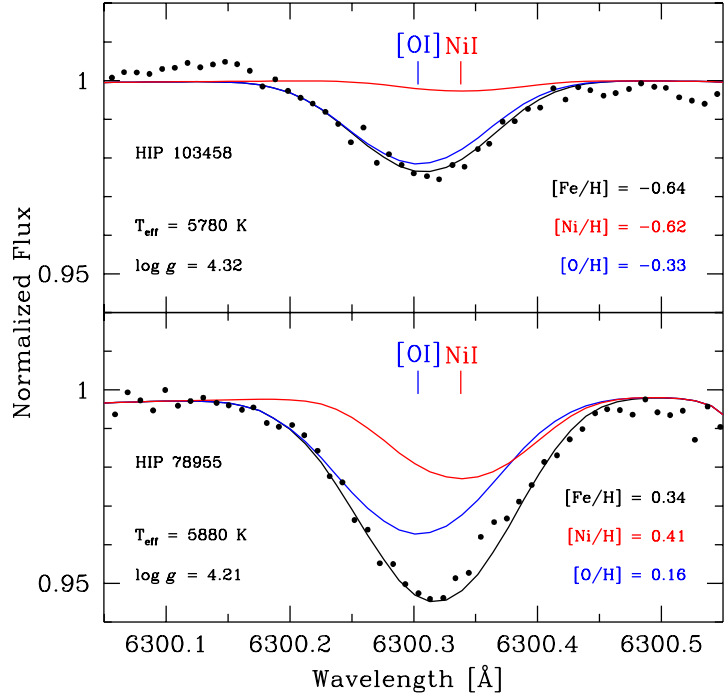}
    \caption{Example of the observed and syntheized spectra for three stars at
             different metallicities. The observed spectra have been shifted in wavelength.
             The fitted oxygen line is marked in blue and the nickel line in red.
             Just to the left of the [OI] line for Hip 103458 there is a contribution
             from the sky emission of $\rm O_{2}$.}
    \label{fig:syntet}
  \end{figure}
  \begin{figure}
    \centering
    \includegraphics[width=7.5cm,angle=-90]{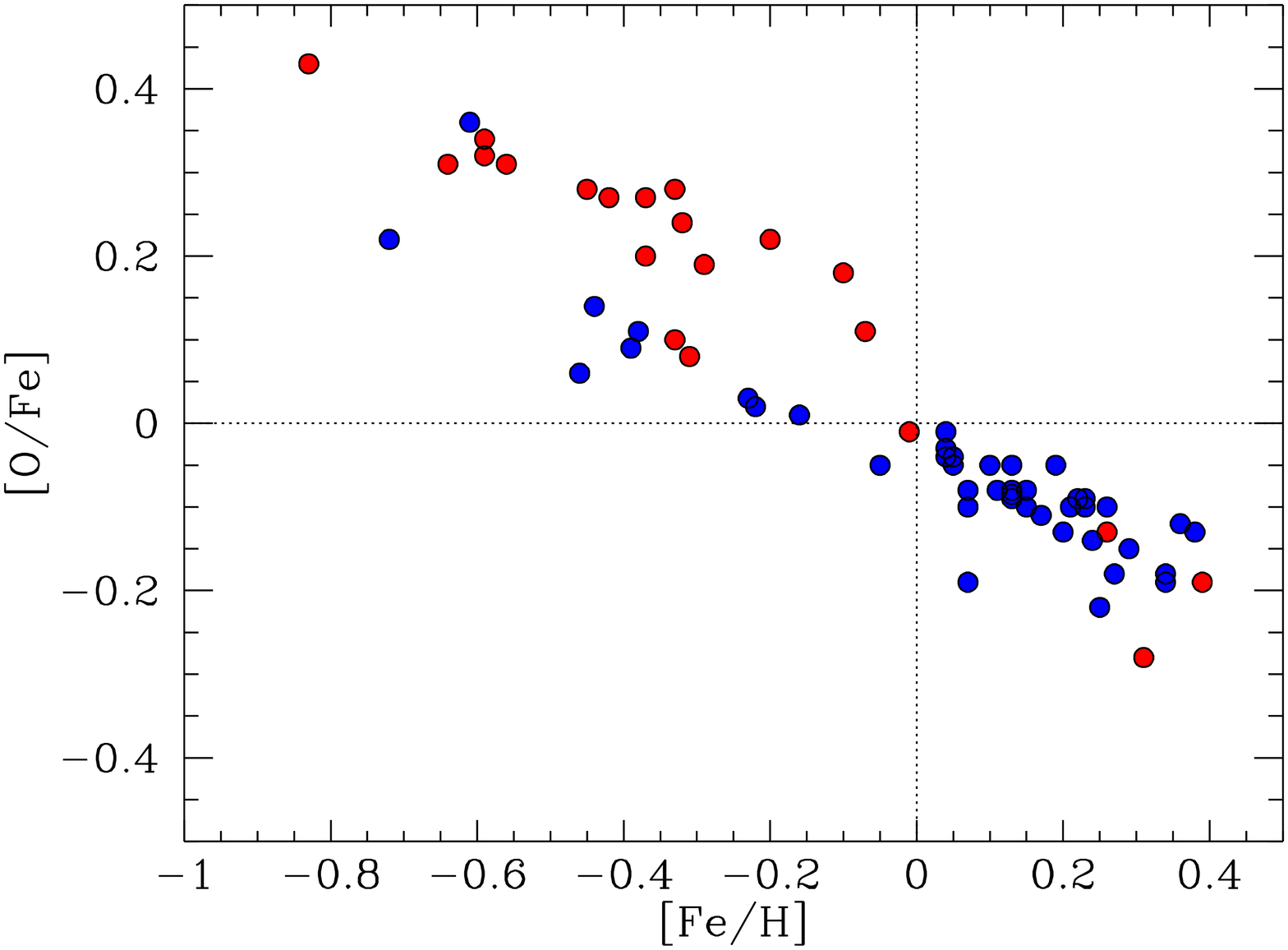}
    \includegraphics[width=7.5cm,angle=-90]{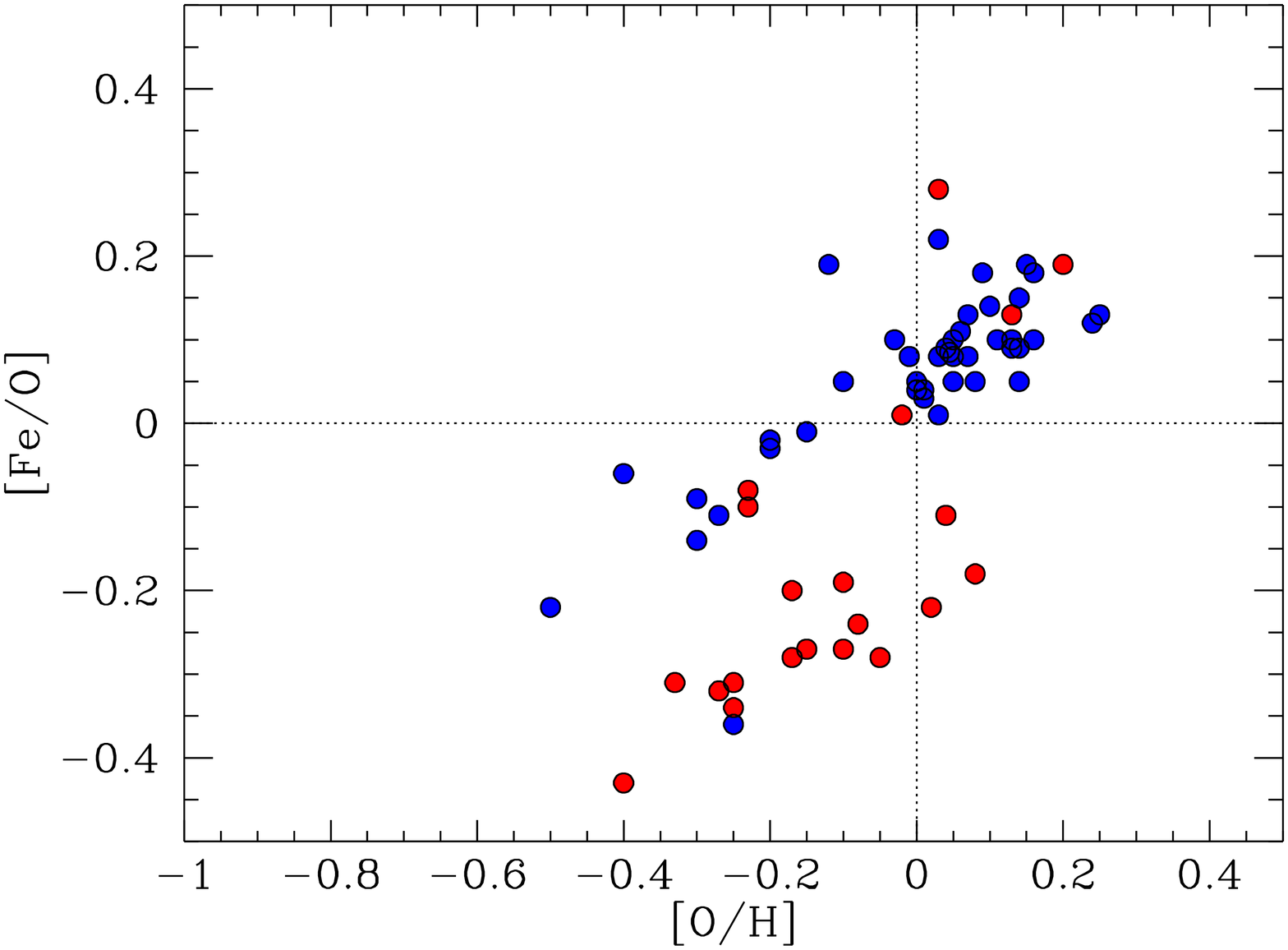}
    \caption{Abundance trends for oxygen.
           Thick disk stars are marked as red, and thin disk stars as blue.}
    \label{fig:syretrender}
  \end{figure}
  \begin{figure}
    \centering
    \includegraphics[width=5.5cm,angle=0]{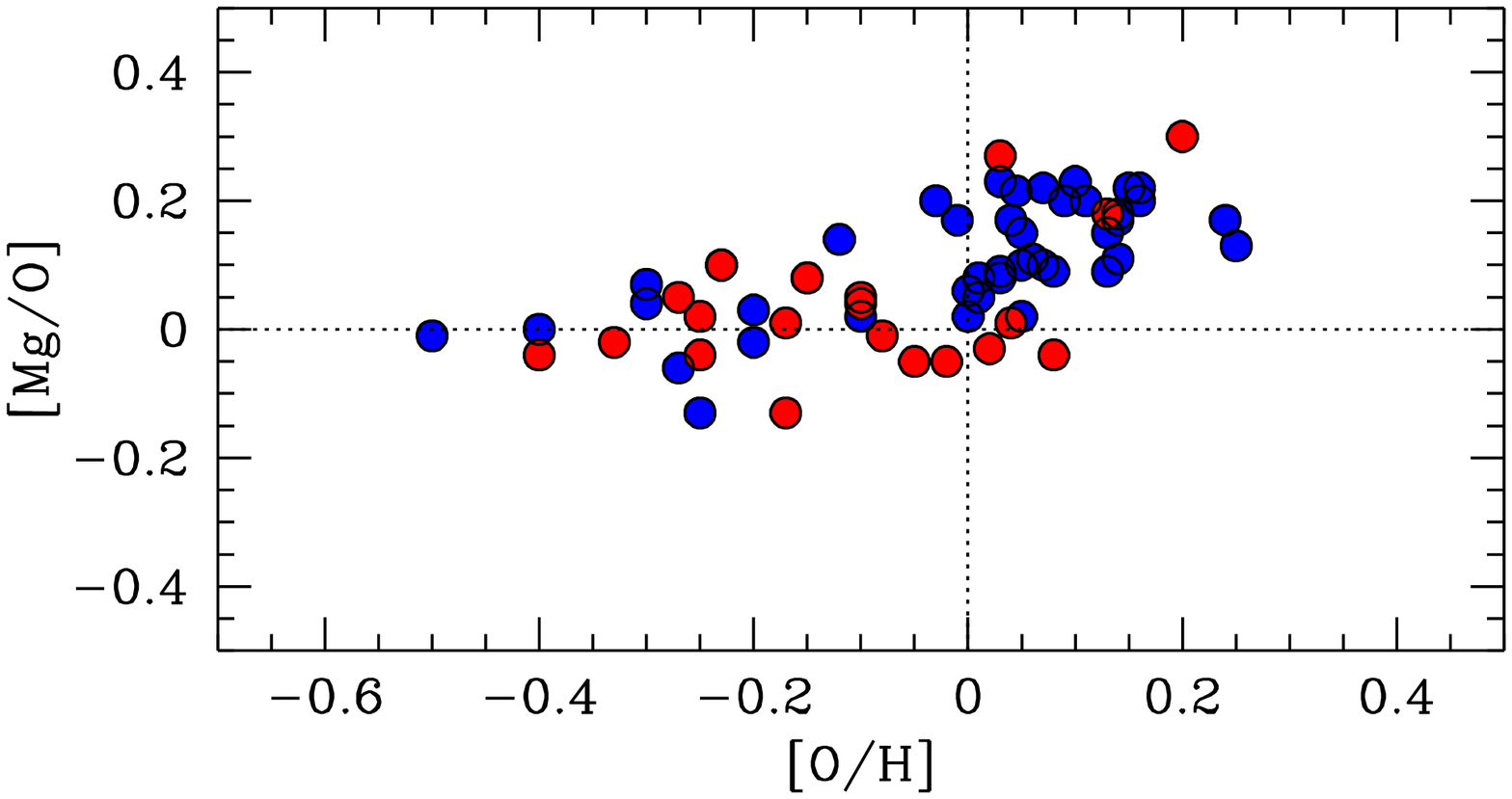}
    \includegraphics[width=5.5cm,angle=0]{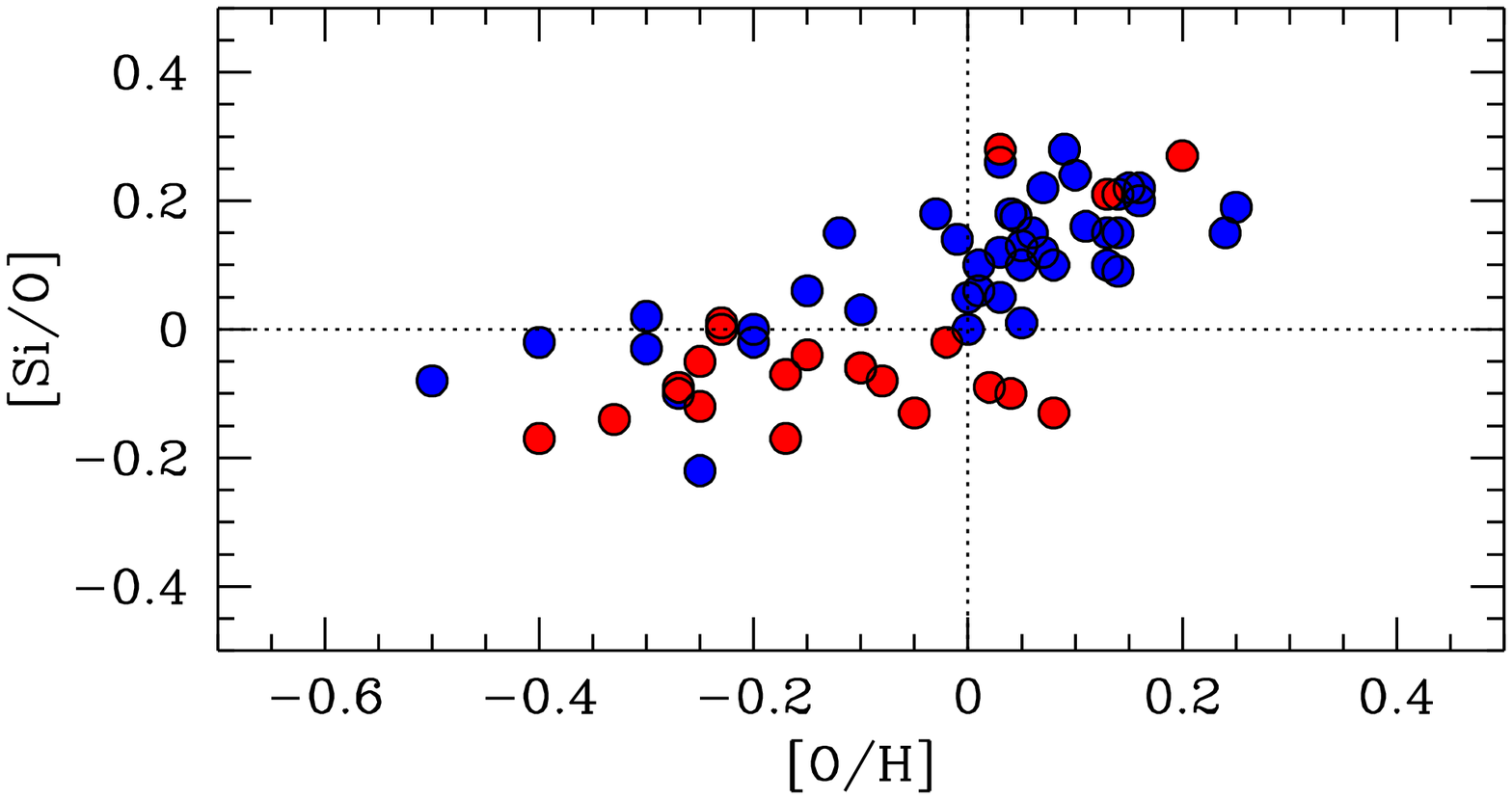}
    \includegraphics[width=5.5cm,angle=0]{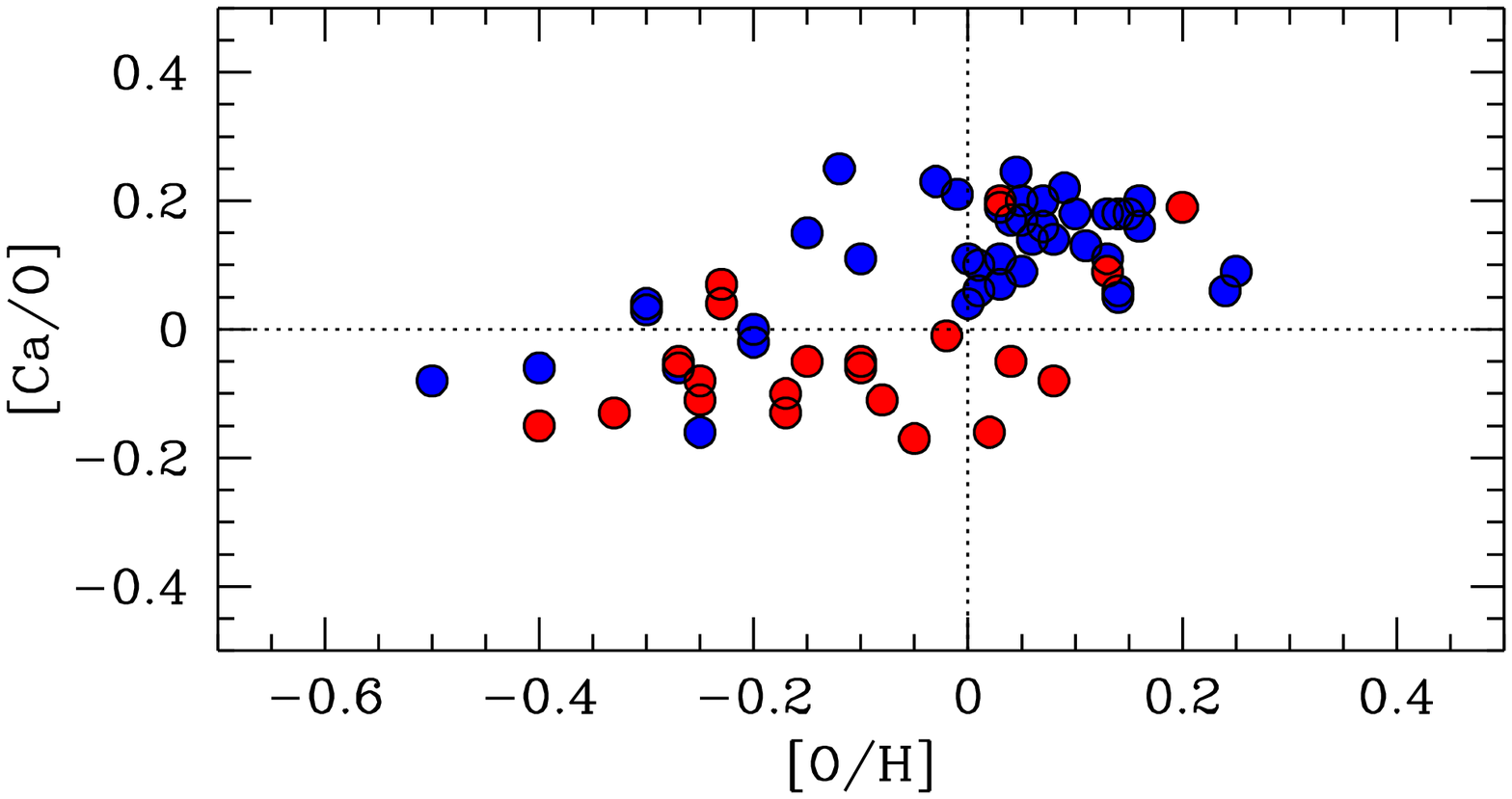}
    \includegraphics[width=5.5cm,angle=0]{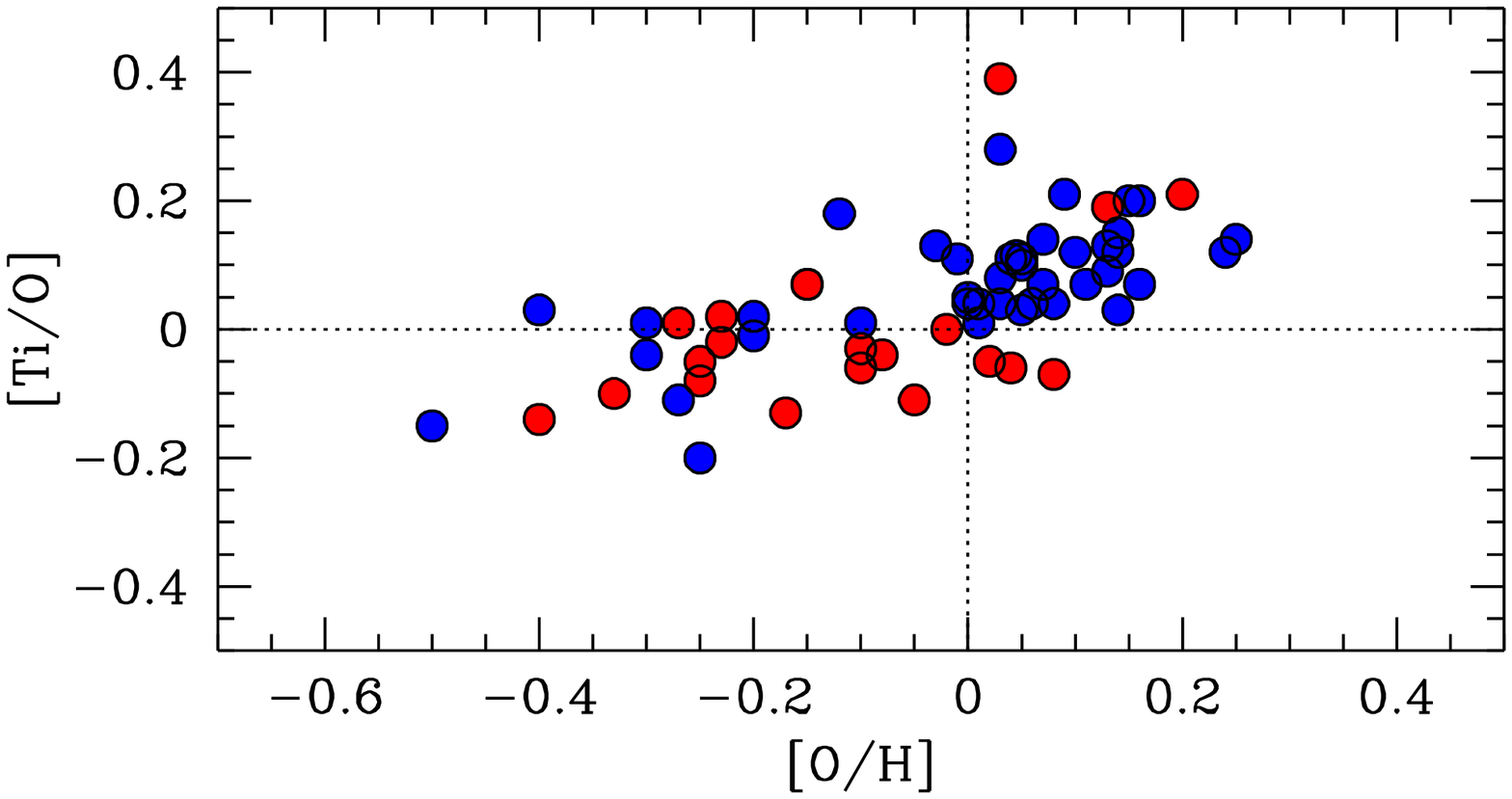}
    \caption{Mg, Si, Ca, and Ti abundances compared to oxygen.
           Thick disk stars are marked as red, and thin disk stars as blue.}
    \label{fig:alpha}
  \end{figure}

\section{Introduction}

The Galactic thin and thick disks are two distinct stellar populations in terms of
age distributions and kinematics. The chemical trends in the two systems are also 
most likely different although recent works give conflicting results, see e.g. 
Chen et al.~(2000) and Fuhrmann~(1998). We show that the abundance trends for 
oxygen are different for the thin and thick disks, when traced by kinematically
selected F and G dwarf stars in the solar neighbourhood. It is not only for oxygen
we see distinct trends between the thin and thick disks. Also the abundance 
trends for Al, Na, Cr, Ni, Zn (Bensby et al. 2003a submitted), Mg, Si, Ca, Ti 
(Bensby et al. 2003a submitted, and Feltzing et al. 2003), Ba, Eu, and Y 
(Feltzing et al. these proceedings) are clearly different for the two disks.

\section{Selection and observation of the stellar sample}

The selection of thin and thick disk stars was based on kinematics and is fully 
described in Bensby et al.~(2003a, submitted). We calculated Gaussian 
probabilities for each star, in a larger sample of $\sim$14\,000 stars, that it 
belongs to the thin and thick disk respectively, using the galactic velocity 
components $U_{\rm LSR}$, $V_{\rm LSR}$, and $W_{\rm LSR}$ of the stars. Stars 
with high probabilities of belonging to either the thin or the thick disk were 
then selected. The sample consists of 21 thick disk stars and 42 thin disk stars. 
In Fig.~\ref{fig:toomre} we show the kinematical properties of our stellar sample.

Spectra were obtained with the CES spectrograph on the ESO 3.6m telescope with 
a resolution of $R\sim 215\,000$ and a signal-to-noise $S/N > 400$. Telluric 
lines were divided out using spectra from fast rotating B stars. Further details 
are given in Bensby et al.~(2003b, in prep).

\section{Synthesis of the [O{\sc i}] line}

The [OI] line have been synthesized taking the Ni\,{\sc i} blend into account. 
From Fig.~\ref{fig:syntet} it is clear that the Ni line is a main contributor to 
the observed line profile, and that O abundances will be severely overestimated 
in metal-rich stars if it is neglected. Atomic data for these two lines have been 
taken from Allende-Prieto et al.~(2001). The stellar atmospheric parameters 
(effective temperature, surface gravity, and microturbulence) were derived from 
FEROS spectra, Bensby et al.~(2003a submitted), using approximately 150 Fe\,{\sc i} 
and 30 Fe\,{\sc ii} lines. Also the Ni abundances to the combined [O\,{\sc i}]-Ni\,{\sc i} 
line profile were taken from this work, where $\sim 50$ spectral lines were used 
in the Ni abundance determination.

\section{The oxygen fossil record}

The four key findings for oxygen in the thin and thick disks, which can be seen in
Fig.~\ref{fig:syretrender}, are: \\
{\bf 1:} Below $\rm [Fe/H]=0$ the thin and thick disks show distinct and
      smooth abundance trends.
      This indicate that the thin and thick disks formed at epochs that are separated in
      time and in space, and from interstellar gas that were reasonably homogeneous.\\
{\bf 2:}  The presence of SNe type Ia signatures in the thick disk. This shows up as
      a ``knee" in the abundance trend at $\rm [Fe/H] \sim -0.4$, and is interpreted
      as being due to the time delay between the SNe type II and the long-lived SNe type Ia
      in the elemental enrichment of the interstellar medium.
      Star formation must therefore have continued in the thick disk after SNe type Ia
      rate peaked, since we see thick disk stars with $\rm [Fe/H]>-0.4$ that have been
      born out of interstellar gas with a lower $\rm [O/Fe]$.
      This means that the SFR in the thick disk must have initially been fast to
      allow the build-up of oxygen from SNe type II at high metallicities before the
      enrichment from SNe type Ia, producing none or little oxygen, set in. \\
{\bf 3:} Quiet evolution in the thin disk. The abundance trend shows a decline when going from
      overabundances of $\rm [O/Fe] \simeq 0.2$ at $\rm [Fe/H] \simeq -0.7$ and reaching
      $\rm [O/Fe] \simeq -0.2$ at $\rm [Fe/H]=0.4$. That the ``knee" is not present
      in the thin disk implies that the star formation rate in the thin
      disk was quite low, with SNe type II and type Ia going off in an alternative manner.\\
{\bf 4:}  At $\rm [Fe/H]>0$ the linear trend found at sub-solar metallicities continous
      for the thin disk stars.
      In contrast Nissen and Edvardsson~(1992) found the 
      $\rm [O/Fe]$ to level out at these metallicities. However, they did not take the Ni\,{\sc i} 
      blend in the [O\,{\sc i}] line
      into account, which becomes severe at these metallicities, see Fig.~\ref{fig:syntet}.
      The Nissen et al.~(2002) $\rm [O/Fe]$ trend also levels out at $\rm [Fe/H]>0$ even though
      they considered the Ni blend. They did however assume a solar ratio for [Ni/Fe] which 
      according to our Ni results (Bensby et al. 2003a submitted) will lead to an underestimation 
      of the Ni contribution. Our Ni trend show a prominent increase in [Ni/Fe] at $\rm [Fe/H]>0$.

The results we find for oxygen are fully consistent with the results we obtained for
other $\alpha$-elements, Mg, Si, Ca, and Ti, for the same stars from FEROS spectra,
Bensby et al.~(2003a submitted) and Feltzing et al.~(2003).

\section{Implications for nucleosynthesis}

From our FEROS spectra we have derived, using equivalent width measurements, abundances
for the other $\alpha$-elements, Mg, Si, Ca, and Ti (Bensby et al. 2003a submitted).
At $\rm [O/H]<0$ the thick disk trends for these elements are mainly flat when compared to
oxygen. This indicate the common origin of
the $\alpha$-elements, namely from massive stars that become SNe type II. For the thin disk
the trends look different, showing an increasing $\rm [\alpha/O]$ ratio with increasing
 $\rm [O/H]$ and a flat trend at $\rm [O/H]>0$ . This could be due to that SNe type Ia
play a role in the synthesizing of these elements.
Also, the yields from later stellar generations might have become
metallicity dependent?



\begin{thereferences}{}

\bibitem{}
 Allende Prieto, C., Lambert, D.L., \& Asplund, M., 2001, ApJ, 556, L63

\bibitem{}
 Chen, Y.Q., Nissen, P.E., Zhao, G., Zhang, H.W., \& Benoni, T.,
 2000, A\&AS, 141, 491

\bibitem{}
 Feltzing, S., Bensby, T., \& Lundstr\"om, I., 2003, A\&A, 397, L1

\bibitem{}
 Fuhrmann, K., 1998, A\&A, 338, 161

\bibitem{}
 Nissen \& Edvardsson, B., 1992, A\&A, 261, 255

\bibitem{}
 Nissen, P.E., Primas, F., Asplund, M., \& Lambert, D.L., 2002, A\&A, 390, 235

\end{thereferences}
\end{document}